\documentclass[aps,twocolumn,showpacs,lengthcheck,preprintnumbers]{revtex4-1}
\usepackage{amsfonts}
\usepackage{amsmath}
\usepackage{amssymb}
\usepackage{graphicx}
\usepackage{fontenc}
\usepackage{color}
\usepackage{textcomp}
\usepackage[unicode]{hyperref}
\hypersetup{
	pdftitle={lattice},
	pdfauthor={},
	pdfsubject={gap_solitons},
	colorlinks=true,
	linkcolor=blue,
	citecolor=blue,
	filecolor=black,
	urlcolor=blue
}

\begin{document}
\title{Gap solitons and nonlinear Bloch states in Bose-Einstein condensates with current-dependent interactions }

\author{Jintao Xu}
\affiliation{
	School of Science, Xi’an University of Posts and Telecommunications, Xi’an, China
}	
\author{Qian Jia}
\affiliation{
	School of Science, Xi’an University of Posts and Telecommunications, Xi’an, China
}	
\author{Haibo Qiu}
\email{phyqiu@gmail.com}
\affiliation{
	School of Science, Xi’an University of Posts and Telecommunications, Xi’an, China
}
\author{Antonio Mu\~{n}oz Mateo}
\email{ammateo@ull.edu.es}
\affiliation{Departamento de F\'isica, Universidad de La Laguna, La Laguna, Tenerife, Spain}


\begin{abstract} 
	We show how the chiral properties of Bose Einstein condensates subject to current-density interactions and loaded in optical lattices can be observed in the realization of nonlinear Bloch states, whose spectrum lacks the usual periodic structure.
	Chirality is also manifested by spatially localized states, or gap solitons, which are found for positive rotation rates of the lattice at the energy gaps between the linear energy bands, whereas for negative rotations they appear in the semi-infinite gap of the linear spectrum. The stability of extended and localized states is checked through the spectrum of linear excitations and nonlinear time evolution of perturbed states, and the phenomenon of Bloch oscillations is explored. Our results are obtained in quasi 1D ring geometries with feasible experimental parameters. 
\end{abstract}

\maketitle

\section{Introduction}
\label{sec:intro}

 Synthetic gauge fields that depend locally on the density of matter have been recently realized
 in ultracold-atom settings  \cite{Clark2018,Gorg2019,Yao2022,Frolian2022}. The unusual properties of
 these systems were theoretically predicted by means of a non-local unitary transformation that maps the density-dependent gauge into a current-dependent interaction \cite{Aglietti1996,Jackiw1997}. Bose-Einstein condensates (BECs) endowed with such interparticle interactions were shown to exhibit chiral properties in a free expansion, the onset of persistent currents, or the center of mass oscillations \cite{Edmonds2013,Edmonds2015}; additionally, in the absence of external potential, it was demonstrated that chiral bright solitons can exist only if they move along one (but not the opposite) direction \cite{Aglietti1996,Dingwall2019,Bhat2021,Frolian2022}, and that collisions between them differ significantly from those between regular solitons \cite{Dingwall2018,Qian2022}.

Many aspects of this chiral theory remain unexplored, and very recent experimental realizations \cite{Frolian2022} open new prospects for testing the theoretical predictions. A significant subject that was previously restricted to solid state systems, the matter dynamics in periodic potentials, became also accessible to the field of ultracold gases with the realization of optical lattices \cite{Morsch2006,Schafer2020}. As far as we know, this subject has still not been addressed within the chiral theory.

In this work, we focus on BECs that are loaded in optical lattices and subject to interactions that depend locally on the current density.
 The lattice is assumed to be imprinted on a quasi-1D ring, as generated by a tight transverse confinement of the atoms, and to be able to rotate.
In particular, within the framework of a generalized Gross-Pitaevskii equation, we study the properties of nonlinear Bloch waves and gap solitons, and demonstrate their unusual properties. While the energy dispersion of the former states loses the usual periodic structure, and new non-regular Bloch states emerge, the situation of gap solitons within the energy gaps changes drastically with the direction of the rotation rate.
We analyze the  spectrum of linear excitations and show that the stability of both extended and localized states is also conditioned by the chiral properties. Finally, we perform numerical simulations of the equation of motion to explore the stability of stationary states against small perturbations, and also the existence of Bloch oscillations in the presence of current-density interactions.

\section{Model}
\label{sec:Model}
We assume that the condensate wave function $\psi(x,t)$ follows a generalized Gross-Pitaevskii equation in a ring of radius $R$:
\begin{equation}
i\hbar\frac{\partial\psi}{\partial t}=\left[ 
\frac{(-i\hbar\partial_x-m\Omega R)^2}{2m}+U_{\rm latt} +\hbar\kappa J \right]  \psi,
\label{eq:gpe}
\end{equation}
where  $U_{\rm latt}(x)=U_0\,\sin^2(\pi x/d)$ is the lattice potential, with amplitude $U_0$ and lattice spacing $d$, which can rotate with angular velocity $\Omega$. The strength of the current-dependent mean field is measured by the dimentionless parameter $\kappa$, and $J(x,t)=\hbar(\psi^*\partial_x\psi-\psi\partial_x\psi^*)/(i2m)$ is the current density in the lab frame.
 The total number of particles $N=\int dx|\psi|^2$,
and the total energy (which does not include an explicit dependency on the current density)
\begin{equation}
E=\int dx\, \psi^*\left[ \frac{(-i\hbar\partial_x-m\Omega R)^2}{2m}+U_{\rm latt}\right]\psi,
\label{eq:energy}
\end{equation}
are conserved quantities \cite{Aglietti1996}. We particularize our analysis on an $M-$site lattice over the ring, so that  $2\pi R=M\,d$.  As energy reference, we will make use of the lattice recoil energy  $E_L=\hbar^2(\pi/d)^2/(2m)$ \cite{Morsch2006}.

The stationary states take the form $\psi(x,t)=\psi(x)\,\exp(-i\mu t/\hbar)$, with $\mu$ as the energy eigenvalue. In the search of stationary states, it is worth noticing that,
from the continuity equation  $\partial_t|\psi|^2+\partial_x (J-|\psi|^2\,\Omega\, R)=0$, the current density  fulfills $ J-|\psi|^2\,\Omega\, R=J_0$,  where $J_0$ is a constant, thus it 
transforms the equation of motion (\ref{eq:gpe}) into the regular time-independent Gross-Pitaevskii equation
\begin{align}
(\mu-\hbar\kappa\,J_0) \psi=\left[ 
\frac{(-i\hbar\partial_x-m\Omega R)^2}{2m}+U_{\rm latt}+g_\Omega\,|\psi|^2 \right]  \psi, 
\label{eq:gpeff}
\end{align}
where the effective constant-interaction strength is $g_\Omega=\hbar\kappa\Omega R$. Furthermore, if the lattice does not rotate, the initial nonlinear equation (\ref{eq:gpe}) is transformed into the time-independent Schr\"odinger equation 
\begin{align}
	(\mu-\hbar\kappa\,J_0) \psi&=\left[ 
	-\frac{\hbar^2}{2m}\partial_x^2+U_{\rm latt} \right]  \psi, 
	\label{eq:schr}
\end{align}
with the current constraint $J_0= |\psi(x)|^2 \,{\hbar}\partial_x \theta(x)/m $, where $\theta=\arg\,\psi(x)$ is the phase.

 The linear excitations $\delta \psi_j= [u_j,\,v_j]^T$ of stationary states $\psi(x,t)\rightarrow \exp(-i\mu t/\hbar)\,\{\psi(x)+\sum_j [u_j(x)\,\exp(-i\omega_j t)+v_j(x)^*\,\exp(i\omega_j^* t)]\}$, with $j$ being a mode index, can be obtained through the Bogoliubov's equations $B\delta\psi_j=\hbar\omega_j\,\delta\psi_j$. The Bogoliubov matrix is given by
\begin{align}
B=
\begin{pmatrix}
H_{\mbox{\tiny GP}}+i\kappa B_{uu}-\mu & i\kappa B_{uv}\\
i\kappa B^*_{uv}& -H_{\mbox{\tiny GP}}^*+i\kappa B^*_{uu}+\mu
\end{pmatrix},
\label{eq:Bog1}
\end{align}
where $H_{\mbox{\tiny GP}}=(-i\hbar\partial_x-m\Omega R)^2/2m+U_{\rm latt}+\hbar \kappa J$ is the Hamiltonian operator in Eq. (\ref{eq:gpe}), $B_{uu}=\hbar^2(\psi{\partial_x \psi^*}-|\psi|^2{\partial_x})/2m$, and $B_{uv}=-\hbar^2(\psi\partial_x \psi-\psi^2{\partial_x})/2m$.
Linear excitations with complex frequencies, $\Im(\omega_j)\neq 0$, lead to the exponential growth (in the linear regime) of small perturbations on the stationary state that can produce its decay during time evolution.

\begin{figure}[tb]
	\flushleft ({\bf a})\\
	\includegraphics[width=0.9\linewidth]{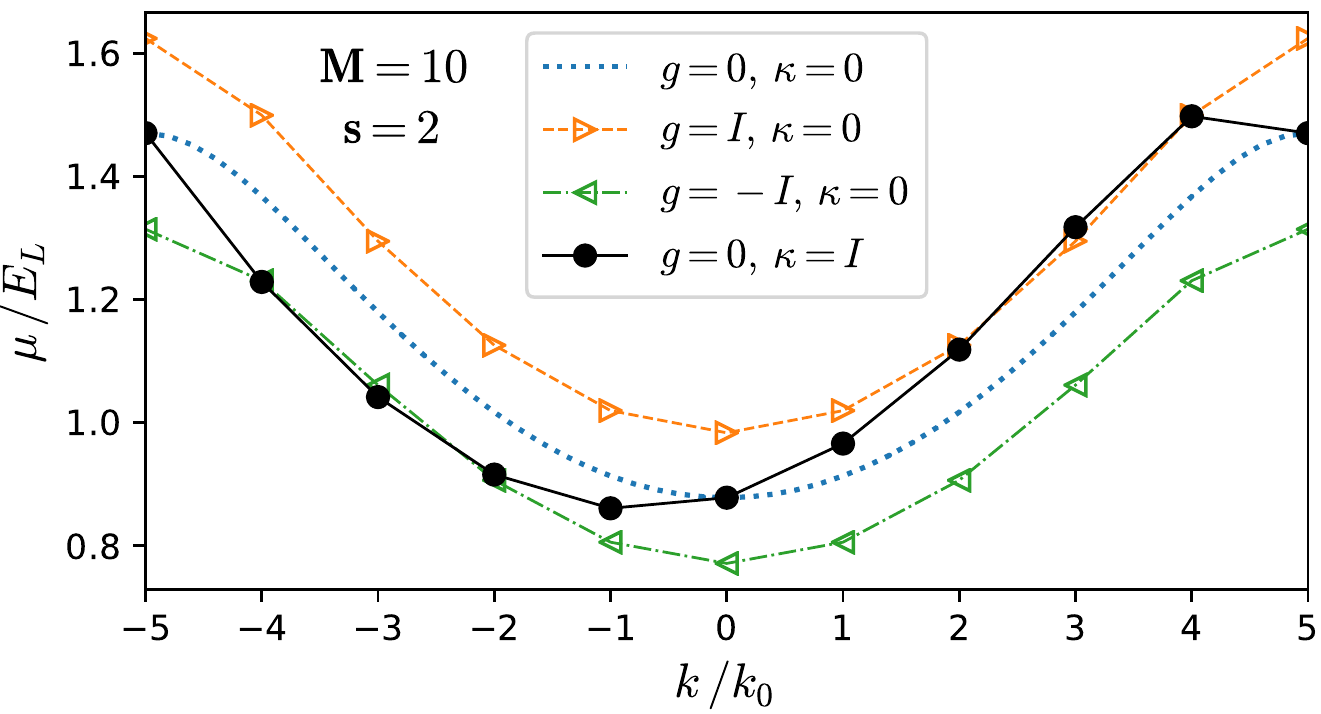}\\ \vspace{-0.3cm}
	({\bf b})\\ 
	\includegraphics[width=0.9\linewidth]{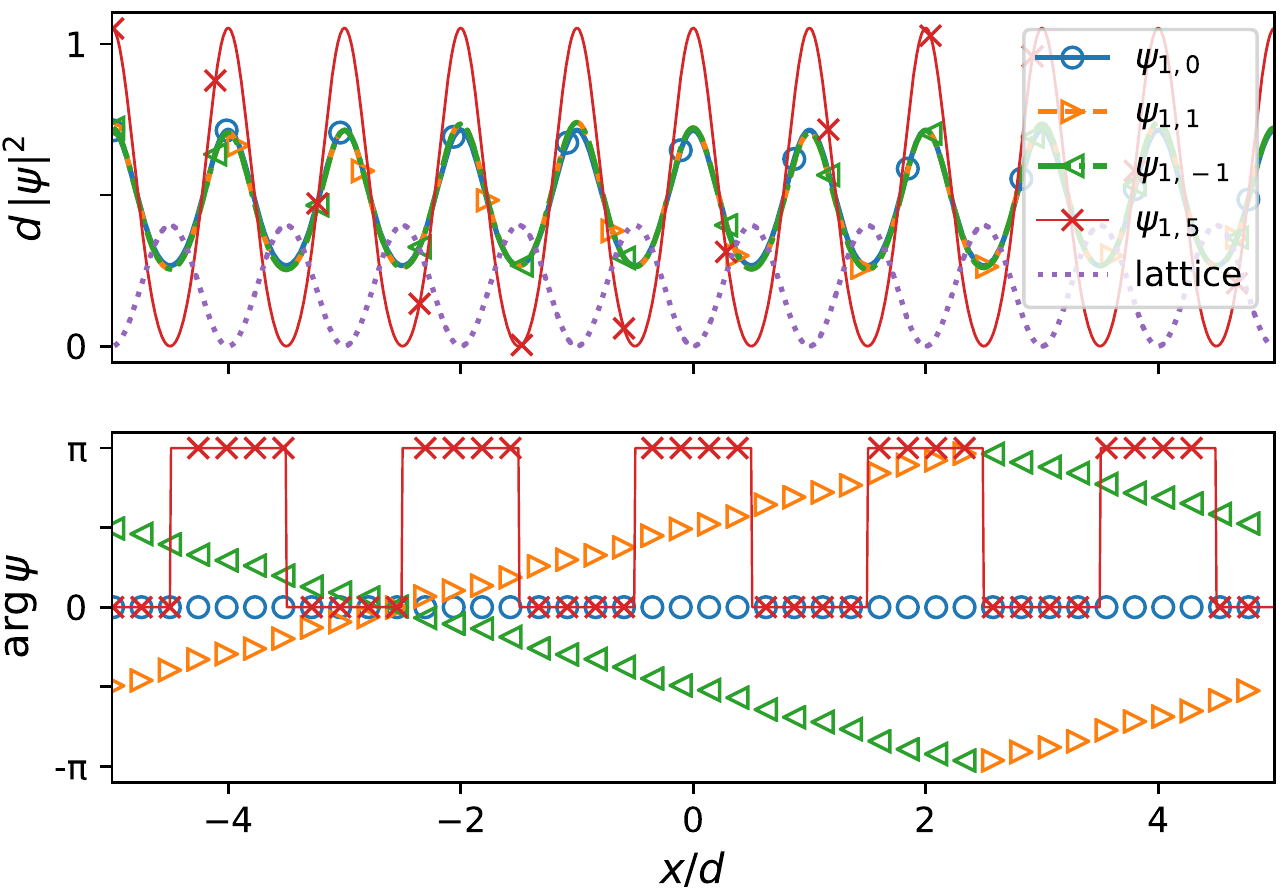}\\ \vspace{-0.3cm}
	({\bf c})\\ 
	\includegraphics[width=0.9\linewidth]{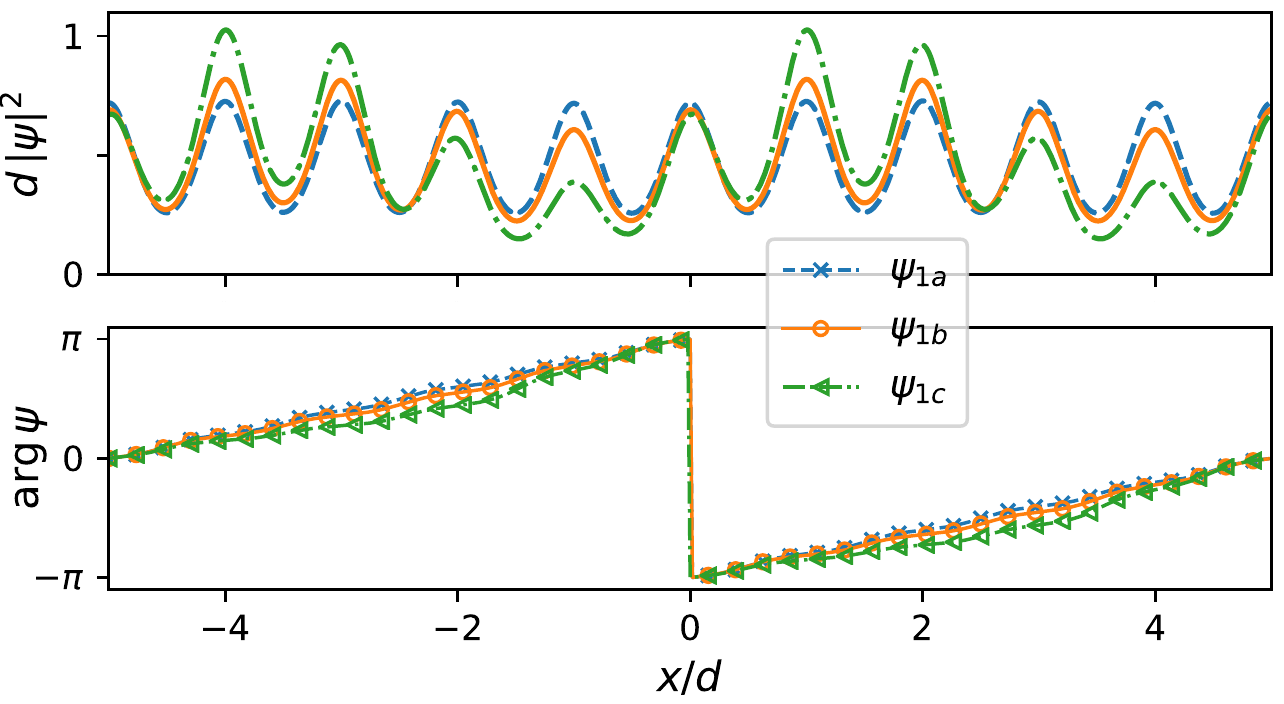}
	\caption{(a) Linear (dotted line) and nonlinear (symbols) lowest energy bands for a ring lattice at rest, $\Omega=0$, with ten sites, $M=10$, lattice depth $s=2$, and average number density fixed by $\kappa\,N/(2\pi R)=0.47/d$. Two types of interparticle interactions with equal strength are represented: contact interaction parameterized by $|g|$ (see text), both repulsive $g>0$ and attractive $g<0$, and current-dependent interaction $\kappa$, with $|g|=\kappa=4.7/N$ (denoted by $I$ in the labels). 
		(b) Density (top) and phase (bottom) profiles of nonlinear Bloch states $\psi_{n,k}$ in the presence of current-density interactions, $\kappa\neq 0$ and $g=0$.
	(c) Three instances of non-regular Bloch states with quasimomentum $k=k_0$, from almost homogenous, $\psi_{1a}$, to large variations, $\psi_{1c}$, in the density peaks.  }
	\label{fig:bands}
\end{figure}

\section{Extendend and localized eigenstates}

In a non-rotating linear system ($\Omega=0$ and $\kappa=0$),
the dispersion relations of quantum states in a ring lattice consist of energy bands separated by energy gaps \cite{Kittel}.
The corresponding spectrum of eigenstates can be described in terms of Bloch waves $\psi_{n,k}(x,t)=\exp[i(kx-\epsilon_{n,k} t/\hbar)]\,u_{n,k}(x)$, with eigenenergies $\epsilon_{n,k}$, where $n=1,\,2\,,\dots$ identifies the band number, and $k=q\,k_0$, with $q=0,\,\pm 1,\,\pm2,\dots$ and $k_0=1/R$,  is the wavenumber associated with the quasimomentum $\hbar k$. The functions $u_{n,k}(x)$ share spatial period with the lattice $u_{n,k}(x+d)=u_{n,k}(x)$, so that the probability density profile is homogeneous over the lattice sites. If the lattice is finite, and contains $M$ sites, there are just $M$ values of quasimomentum \cite{Kittel}. 

In a system with varying contact interactions, the linear Bloch waves have been shown to find continuation as nonlinear Bloch waves when the interactions are switched on \cite{Morsch2006, Mateo2019}. We will show that this continuation also exists in the presence of current-density interactions. In addition, differently to the case of contact interactions, there exist new extended states with a non-homogeneous density profile over the lattice sites.

\subsection{Nonlinear Bloch waves}
First, we focus on the dispersion of the system at $\Omega=0$ for varying quasimomentum. Insight can be obtained from the comparison with a system subject to contact interactions (hence following the usual GP equation); in this case, the linear energy bands are shifted to higher energies when the interaction is repulsive, whereas the opposite happens for attractive interaction. Therefore, in the presence of current-density interaction, where the effective interaction changes from repulsive to attractive according to the sign of the particle current, the resulting dispersion curves are expected to be asymmetric with respect to the value of quasimomentum, with energies higher than the linear bands for states with positive currents, and lower than the linear bands for states with negative currents.

As can be seen in Fig. \ref{fig:bands}(a), this is indeed the scenario shown by our numerical results for a ring lattice with $M=10$ sites and shallow depth $s=2$. The number of particles has been fixed for the nonlinear states considered to give $\kappa N/(2\pi R)=0.47/d$; correspondingly, the order of magnitude of the nonlinearity in Eq. (\ref{eq:gpeff}) is 
$g_\Omega N/ (2\pi R)= 4.7 (R/d) \hbar\Omega$.
For comparison, contact interaction cases are represented, and have been parameterized by the non-dimensional quantity $g=m \,d\,g_{\rm 1D}/\hbar^2 $, where $g_{\rm 1D}$ is the one-dimensional contact interaction strength, so that $|g|=\kappa$. The lowest energy band of the linear system  (dotted line) lays in between the lowest chemical potential bands of nonlinear systems with positive (open symbols joined by dashed lines to guide the eye) and negative (open symbols joined by dot-dashed lines) contact interactions, whereas the energy eigenvalues of the system with current-density interaction (filled symbols) vary as predicted. The states with minimum (zero) and maximum ($k=5\,k_0$) values of quasimomentum have no particle currents, thus they follow a linear (Schr\"odinger) equation of motion, and match the energy of the linear bands.

Four instances of Bloch states, with quasimomentum wavenumber $k/k_0=1,\,-1,\,5,$ represented by their density and phase profiles, are shown in Fig. \ref{fig:bands}(b). While states with opposite currents but equal absolute value of quasimomentum $|k|$ do not show appreciable differences in the density profile, states with different $|k|$ do, reflecting the increasing interaction associated to higher $|k|$.  

\subsection{Non-regular Bloch states}

An interesting novelty of the systems with current-density interactions, contrary to the case of contact interactions, is the existence of extended stationary states that do not conform to the usual picture of Bloch states, since they present a non-homogeneous density profile over the lattice sites. They do not conform either to the features of alternative states hosting dark solitons in the lattice \cite{Machholm2004}. We will refer to them as non-regular Bloch states, since the quasimomentum, associated with the phase winding number $q=k/k_0$, is still a well defined quantity. As a general aspect, the higher the variation between density peaks of lattice sites that they present, the lower the constant current density becomes. Our numerical results for characteristic quantities suggest a continuum of non-regular Bloch states, for we were able to find close states with very small differences, of the order of 1$\mbox{\textperthousand}$ in energies. The linear Eq. (\ref{eq:schr}) sheds light on this phenomenon, since it admits solutions as linear superpositions of states with the same eigenvalue $\mu-\hbar\kappa\,J_0$ and current density $J_0$; such superpositions are possible because the linear Bloch states are doubly degenerate.

Figure \ref{fig:bands}(c) shows the density (top) and phase (bottom) profiles of three non-regular Bloch states with winding number $q=k/k_0=1$. They range from almost homogeneous $\psi_{1a}$, to intermediate $\psi_{1b}$, and up to large variation $\psi_{1c}$, in the density peaks. The density modulation over the whole lattice has the form $n_k\,[1+\beta\,\sin(k_0\,x)]$, where $n_k$ is the density of the homogenous Bloch state, and $\beta$ varies from almost zero for $\psi_{1a}$, to $\beta=0.6$ for $\psi_{1c}$. Despite the large differences in the density profile, their energy eigenvalues differ in less than 1$\%$, and their energies in less that 1$\mbox{\textperthousand}$. The phase profiles show also small differences and follow a monotonic increase in the range $[0,\,2\pi]$. 

\begin{figure}[tb]
	\includegraphics[width=\linewidth]{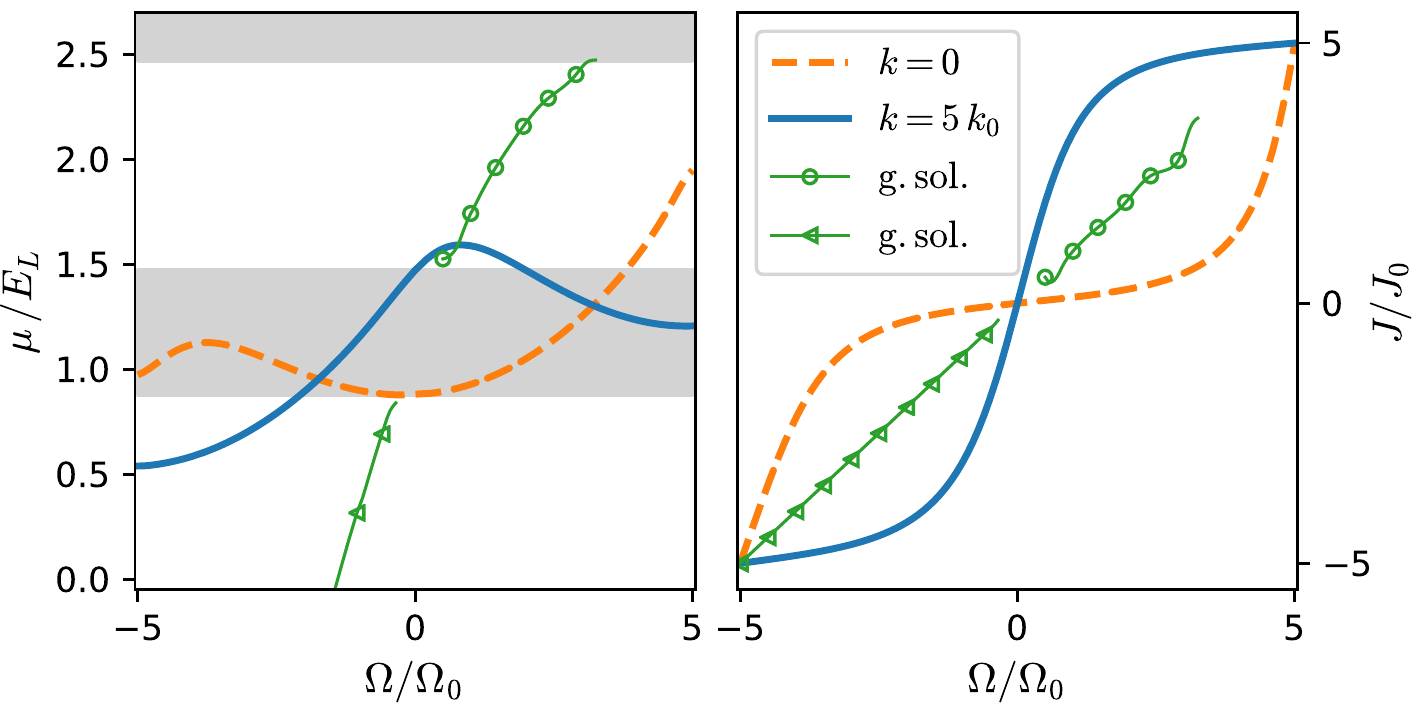}
	\caption{ Trajectories of nonlinear Bloch waves (solid and dashed lines) and gap solitons (thin lines with symbols)  in a ring lattice moving with angular rotation $\Omega$. All states contain the same number of particles, so that the varying rotation translates into a varying interaction. Nonlinear Bloch waves are characterized by the wavenumber $k$ that indexes their quasimomentum, whereas gap solitons trajectories differ for positive (open circles) and negative (open triangles) currents.  Left: Energy eigenstates measured in units of the lattice recoil energy $E_L$; the underlying energy bands of the linear problem (shaded regions) are represented for comparison. Right: Average current density in units of $J_0=N\Omega_0$.
	}
	\label{fig:traj}
\end{figure}
\begin{figure}[tb]
	\flushleft ({\bf a})\\
	\includegraphics[width=\linewidth]{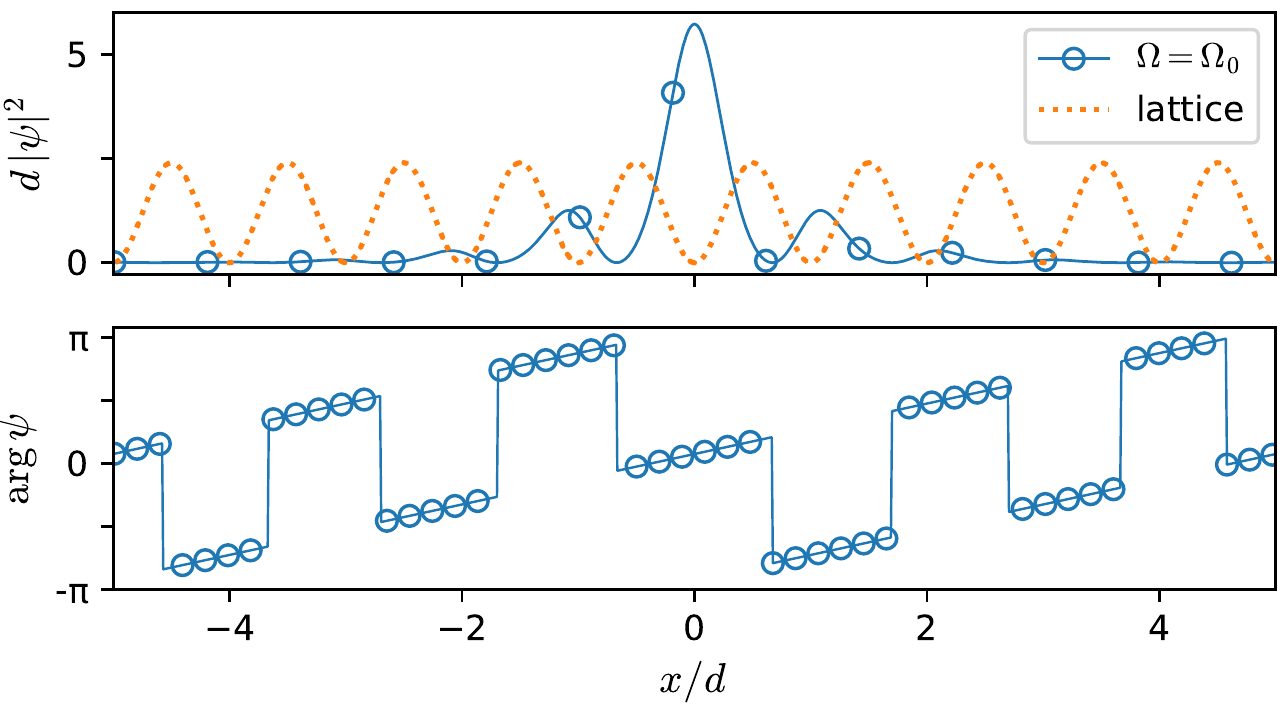}\\ \vspace{-0.4cm}
	\includegraphics[width=\linewidth]{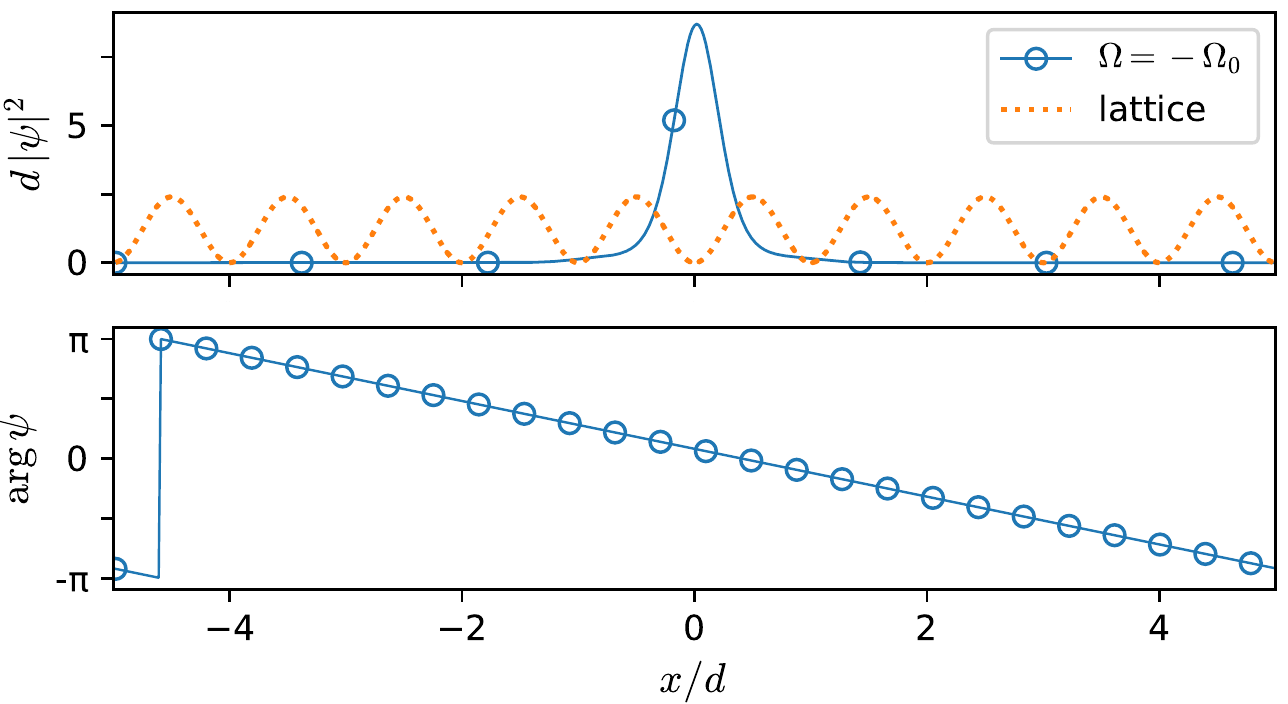}\\ \vspace{-0.4cm}
	({\bf b})\\
	\includegraphics[width=\linewidth]{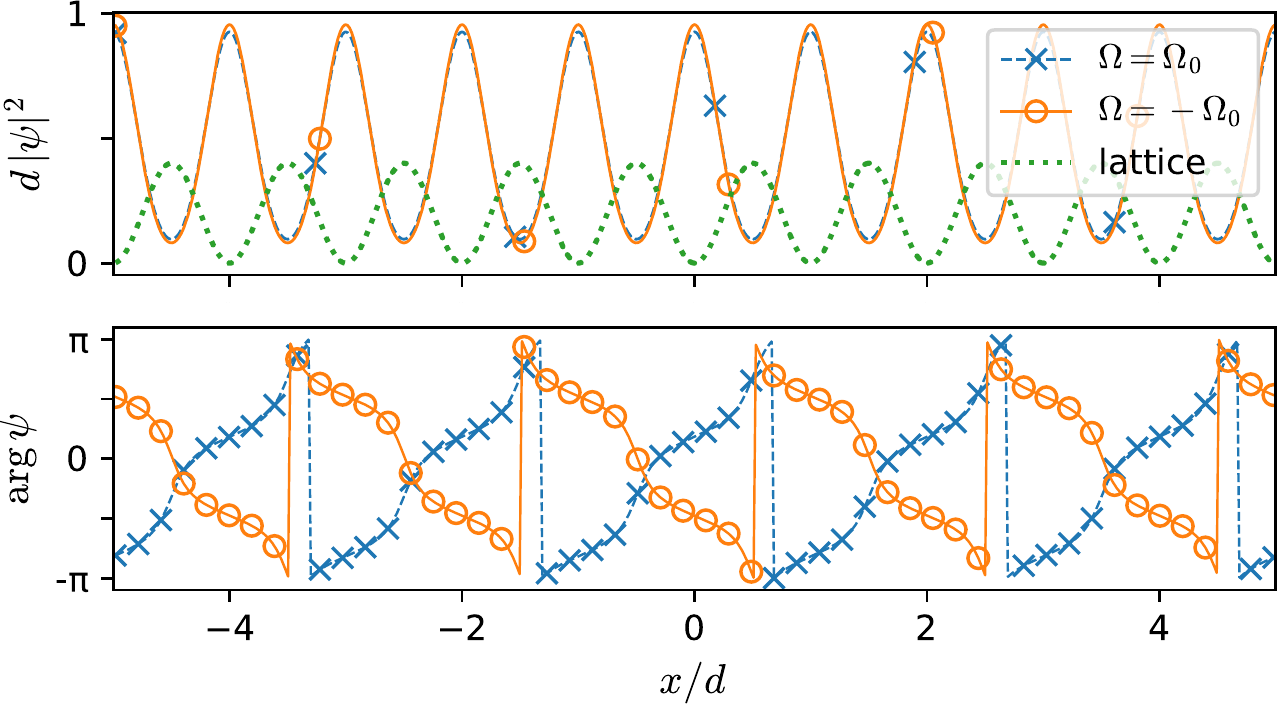}
	\caption{ Gap solitons (a)  and  nonlinear Bloch waves with quasimomentum $k=5\,k_0$ (b), for positive $\Omega=\Omega_0$ and negative $\Omega=-\Omega_0$ lattice rotations, and otherwise equal parameters to those used in Fig. \ref{fig:traj}.
	}
	\label{fig:gs}
\end{figure}

\subsection{Lattice rotation and gap solitons}

Gap solitons are localized states in systems loaded in optical lattices; they usually occupy a few sites, a small region of the whole lattice. Although their existence can be easily understood in systems with attractive interactions, similarly as in translation-invariant settings, their emergence in the presence of repulsive interactions is a priori not that evident \cite{Louis2003}, and can be explained through the sign change induced by the lattice in the effective mass of the particles (see for instance Ref. \cite{Kramer2003}).
In this work, the current-density interaction provides both possibilities for the emergence of gap solitons, which become distinct for positive and  negative current densities. Since the
interaction (or nonlinearity) is necessary for the solitons to exist, the ring lattice has to rotate in order for the gap solitons to emerge.

Although the finite lattice considered here, having $M$ sites, allows for only $M$ Bloch waves, the introduction of rotation gives access to the continuous spectrum of the infinite lattice \cite{Mateo2019}. In systems with Galilean symmetry (as happens for contact interactions), for varying rotation rate $\Omega\in(-M/2,M/2]\times \Omega_0$, where $\Omega_0= \hbar/(mR^2)$, the energy of each Bloch wave $\epsilon_{n,k}(\Omega)$ in the finite lattice [as obtained from Eq.(\ref{eq:energy})] reproduces the energy band profile against quasimomentum  in the first Brillouin zone $k\in(-\pi/d,\pi/d]$ of the infinite lattice with $\Omega=0$. The dipersion graph, $\epsilon_{n,k}$ versus $\Omega$, is also useful in understanding the emergence of gap solitons; the energy degeneracies found in this graph for the linear system, which  correspond to crossings of Bloch-wave trajectories, provide the origin of gap solitons when the interparticle interactions are switched on. Thus, gap solitons are the nonlinear continuation of linear states made of Bloch-wave superpositions \cite{Mateo2019}.

Figure \ref{fig:traj} shows our numerical results for stationary states in a moving ring lattice with current density interactions and same parameters as in Fig. \ref{fig:bands}. The eigenenergy (left) and current density (right) of both nonlinear Bloch states and gap solitons are represented for fixed number of particles.
Gap solitons spread when approaching the energy bands (light-gray shaded regions in the graph), and then become dynamically unstable \cite{Louis2003}. Eventually, as happens in the present case at the bottom of the second energy band, they extend to the whole system (or stop existing in an infinite lattice) when entering a band \cite{Mateo2019}. Overall,
the chirality of the system manifests as asymmetric trajectories for positive and negative rotation rates of the lattice. In addition, as we demonstrate next, apparent differences arise in the  states belonging to these families, which show distinct density profiles and stability properties.

Figure \ref{fig:gs}(a) shows the density and phase profiles of two typical gap solitons with the same number of particles and opposite lattice rotation. For positive rotation rate (top panel) the soliton is situated between the first and second energy bands, corresponding to the family represented by lines with open circles in Fig. \ref{fig:traj}. For negative rotation (bottom panel) the soliton belongs to the family lying in the semi-infinite gap, indicated by lines with open triangles in  Fig. \ref{fig:traj}. The latter soliton, having negative current and then effective attractive interparticle interaction, is comparatively more compact than the former, and occupies just one lattice site. On the contrary, as can be seen in Fig. \ref{fig:gs}(b), the density profiles of two nonlinear Bloch waves with equal quasimomentum $k=5\,k_0$ but opposite lattice rotations, hence opposite current densities, are almost indistinguishable (notwithstanding, the differences become clearer for increasing number of particles).

\begin{figure}[htb]
	\flushleft	 ({\bf a})\\ \vspace{-0.25cm}
	\qquad	\includegraphics[width=0.8\linewidth]{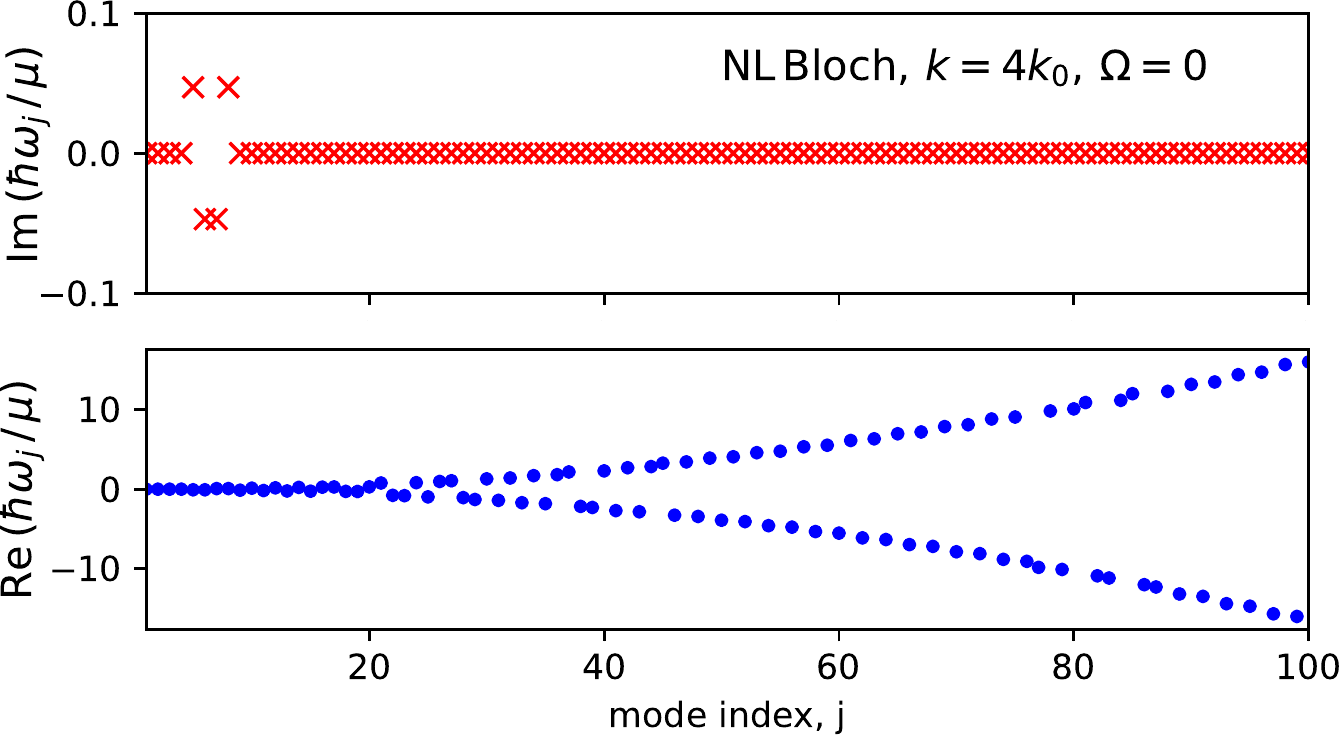}\\
	\qquad\includegraphics[width=0.8\linewidth]{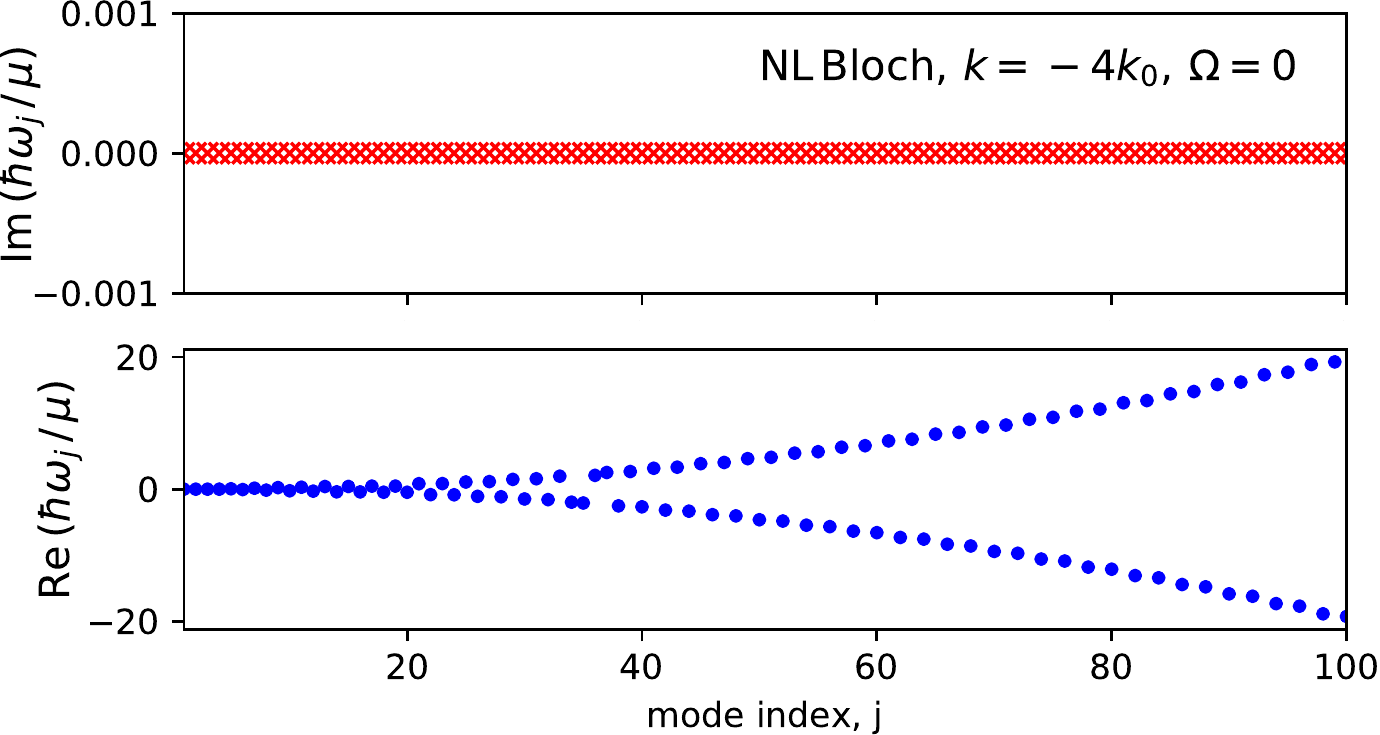}\\ \vspace{-0.25cm}
	\flushleft	({\bf b})\\ \vspace{-0.25cm}
	\qquad	\includegraphics[width=0.8\linewidth]{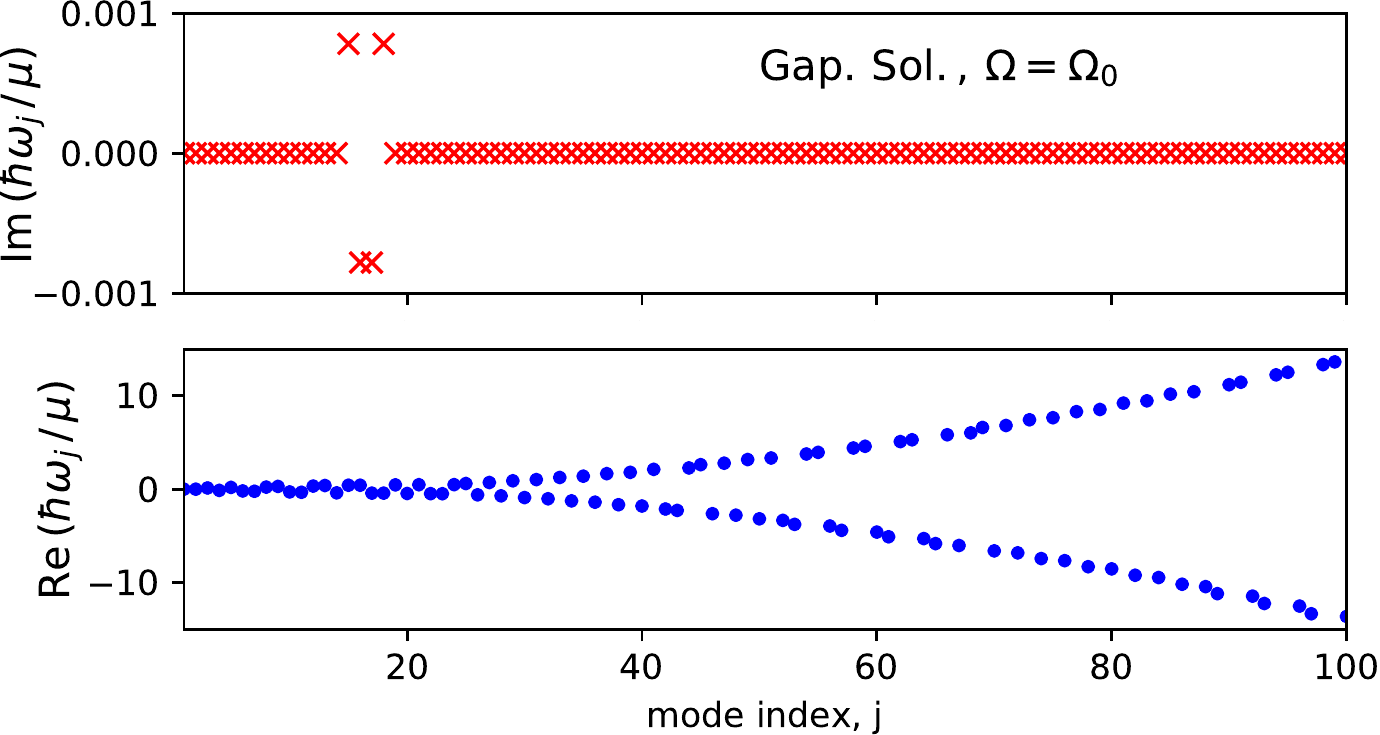}\\
	\qquad	\includegraphics[width=0.8\linewidth]{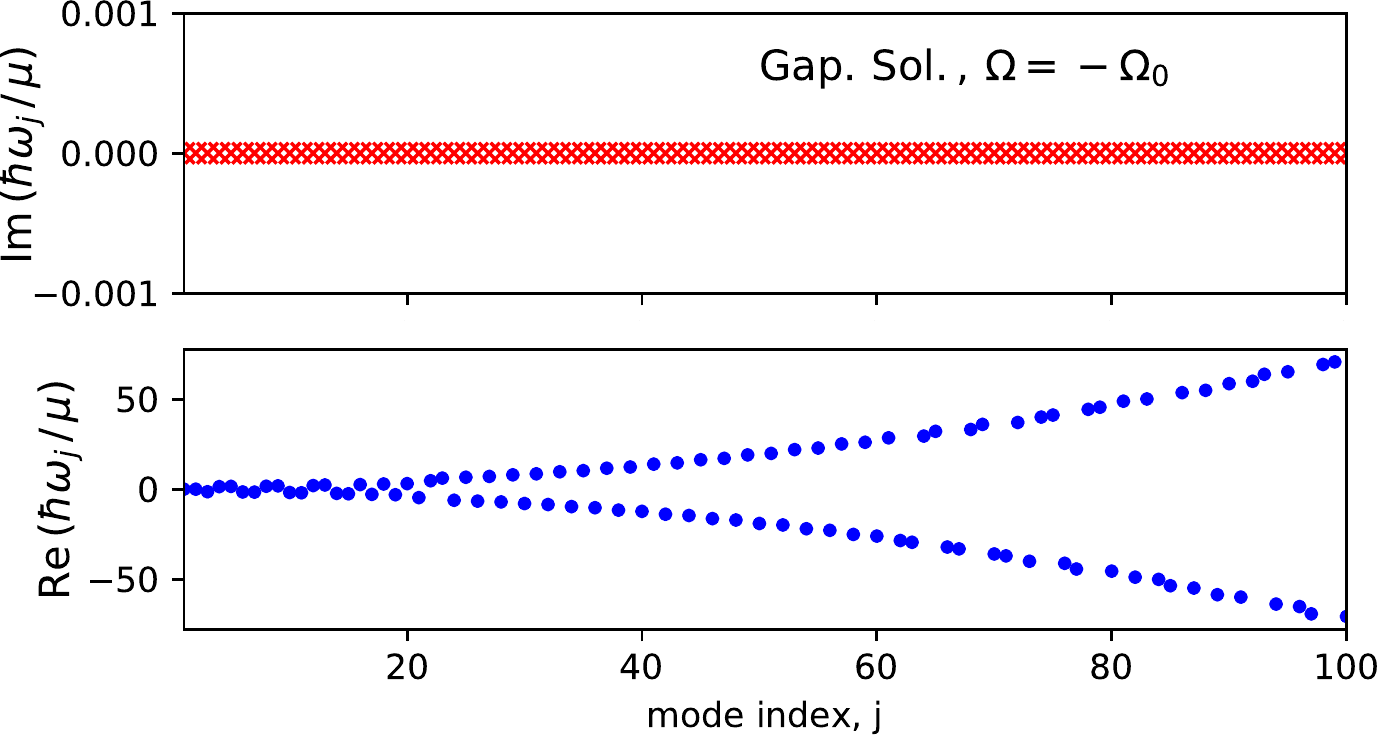}
	\caption{Linear excitation modes of stationary states for fixed number of particles. (a) Nonlinear Bloch states with $k=4\,k_0$ in a lattice at rest.  (b) Gap solitons with  opposite rotation rates shown in Fig. \ref{fig:gs}.   } 
	\label{fig:modes}
\end{figure}

\subsection{Linear stability analysis}
We have studied the linear stability of the stationary states reported in Fig. \ref{fig:traj} by numerically solving the corresponding Bogoliubov's equations (\ref{eq:Bog1}). Before analyzing our results,
it is insightful to remind the scenario of equivalent states with contact interparticle interactions; we particularize it for otherwise equal parameters as in Fig. \ref{fig:traj}. In such a case, while there is no difference regarding the sign of the quasimomentum, the dynamical stability depends strongly on the character of the contact interactions, either repulsive or attractive. For the former case, it is known that Bloch states close to the edge of the Brillouin zone become unstable; how close depends in turn on the strength of the interactions and the lattice depth \cite{Wu03}. The opposite happens for attractive contact interactions, where Bloch states close to zero quasimomentum become unstable; in this latter situation, one can understand the source of instability as associated with smoother, slowly varying density profiles, on which modulation instability can operate due to the existence of lower energy states with localized density profiles. Regarding fundamental (one main peak) gap solitons, althouth instabilities can be found when their chemical potential approaches a linear energy band, they are usually stable states.

Our results in current-interacting systems show, in general, a trend similar to the scenario of contact interactions. The main difference resides in the asymmetry between positive and negative quasimomenta, which can be mapped into systems with effective positive and negative contact interactions, respectively. Another particular difference is observed in the stability of gap solitons with positive rotation rates for the lattice depth considered in Fig. \ref{fig:traj}, $s=2$, for which we have not found stable cases, despite the fact that the energy of the unstable modes (having complex frequencies) can be very small in comparison with the corresponding energy eigenvalue $\mu$. Still, we did find stability for these solitons at higher values of the lattice depth.

Several examples of these general features on linear stability are presented in 
 Fig. \ref{fig:modes};  both the real and imaginary part of the excitation modes are shown for each state considered. 
  The panels (a) show the excitation energies of two nonlinear Bloch waves with the same absolute value of quasimomentum $|k|=4k_0$ (close to the edge of the Brillouin zone) in a lattice at rest  $\Omega=0$. While the state with negative quasimomentum is dynamically stable, the positive-quasimomentum state is not. The scenario is analogous to systems with contact interparticle interactions, hence our results are the opposite (stable for positive rotation and unstable for negative rotation) for states with quasimomentum $|k|=k_0$ (not shown). 
 
 The panels in Fig. \ref{fig:modes}(b) represent the energy excitations of the two gap solitons shown in Fig. \ref{fig:gs}. As anticipated, the soliton with negative rotation is dynamically stable, while the soliton with positive rotation,
  $\Omega=\Omega_0$, presents complex frequencies that can cause the dynamical decay of this stationary state in a real-time evolution (see next section). However, for the same parameters but in a deeper lattice, $s=3$, we have found that the corresponding soliton becomes stable.

\begin{figure}[t]
	\flushleft ({\bf a})\\
	\includegraphics[width=1.0\linewidth]{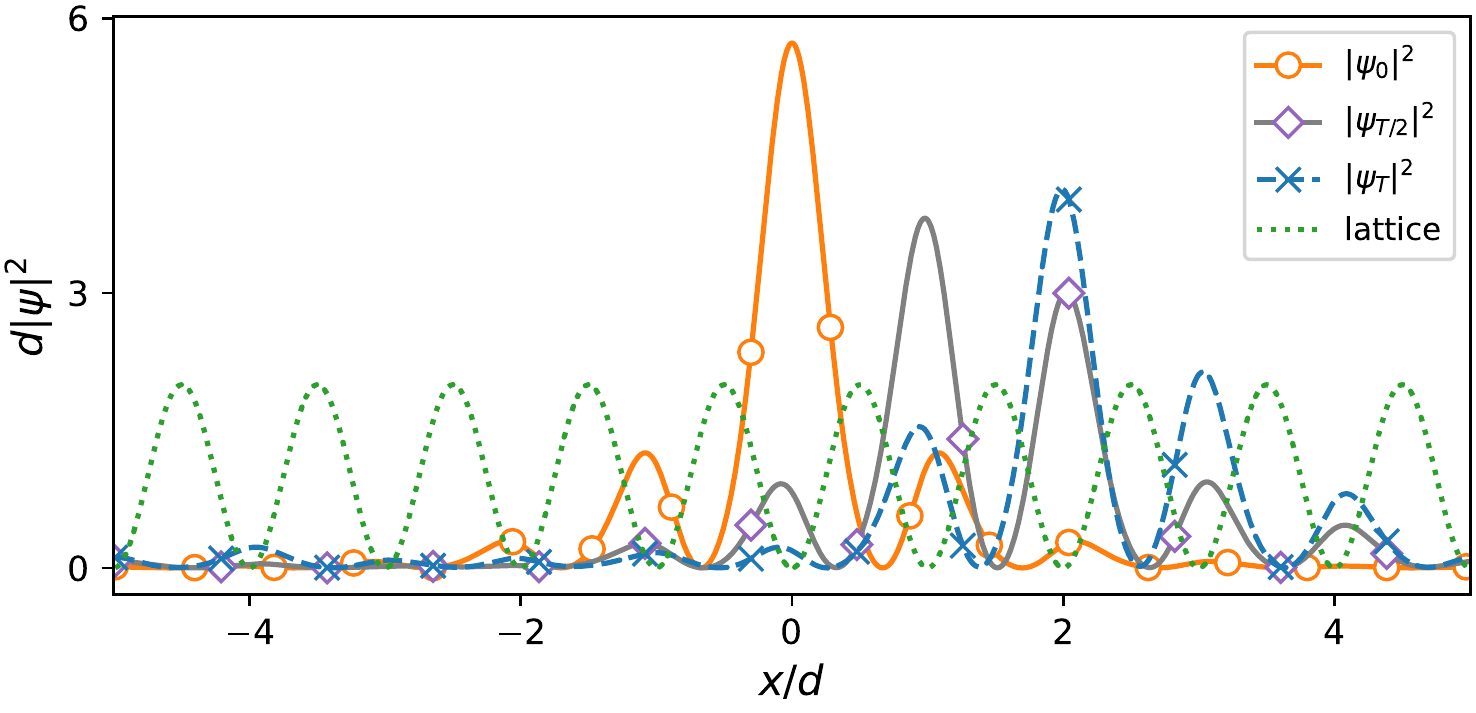}\\
	({\bf b})\\
	\includegraphics[width=1.0\linewidth]{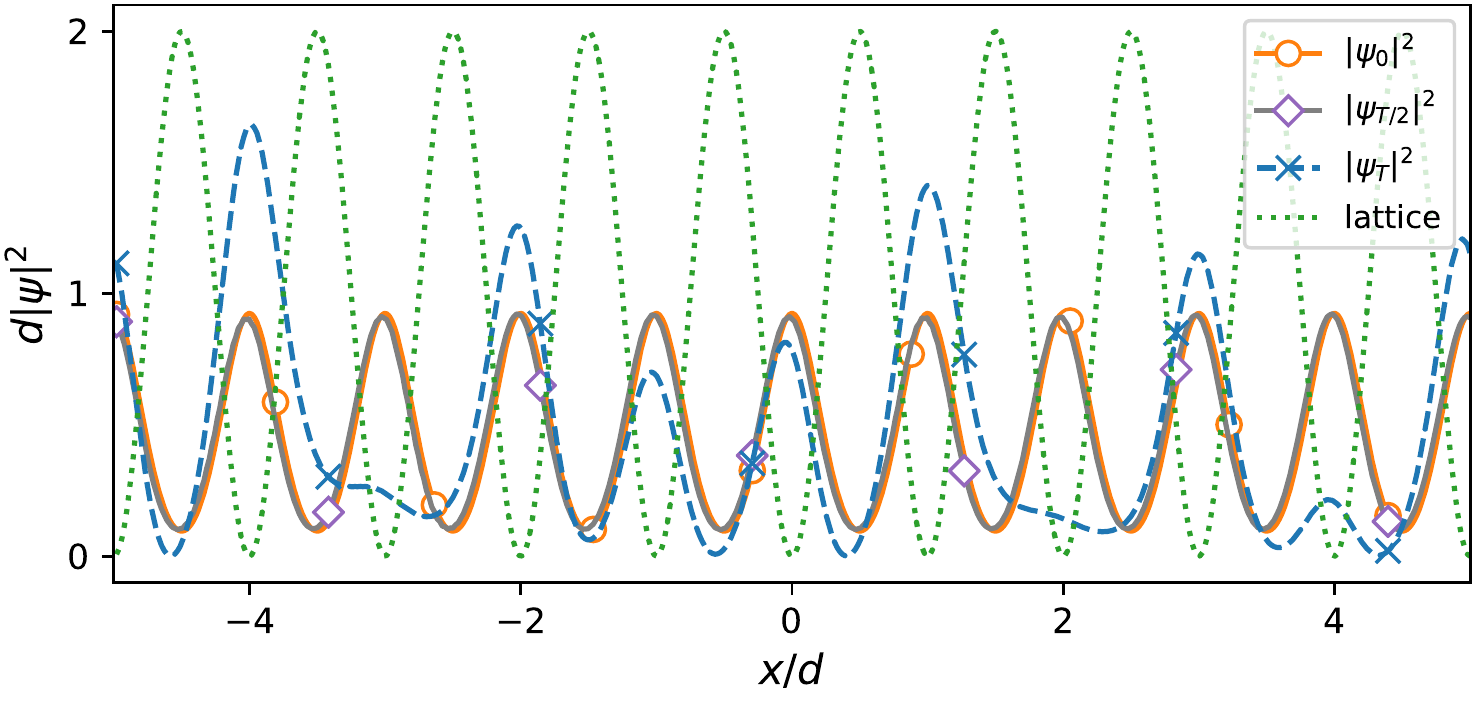}\\ \vspace{-0.4cm}
	\includegraphics[width=1.0\linewidth]{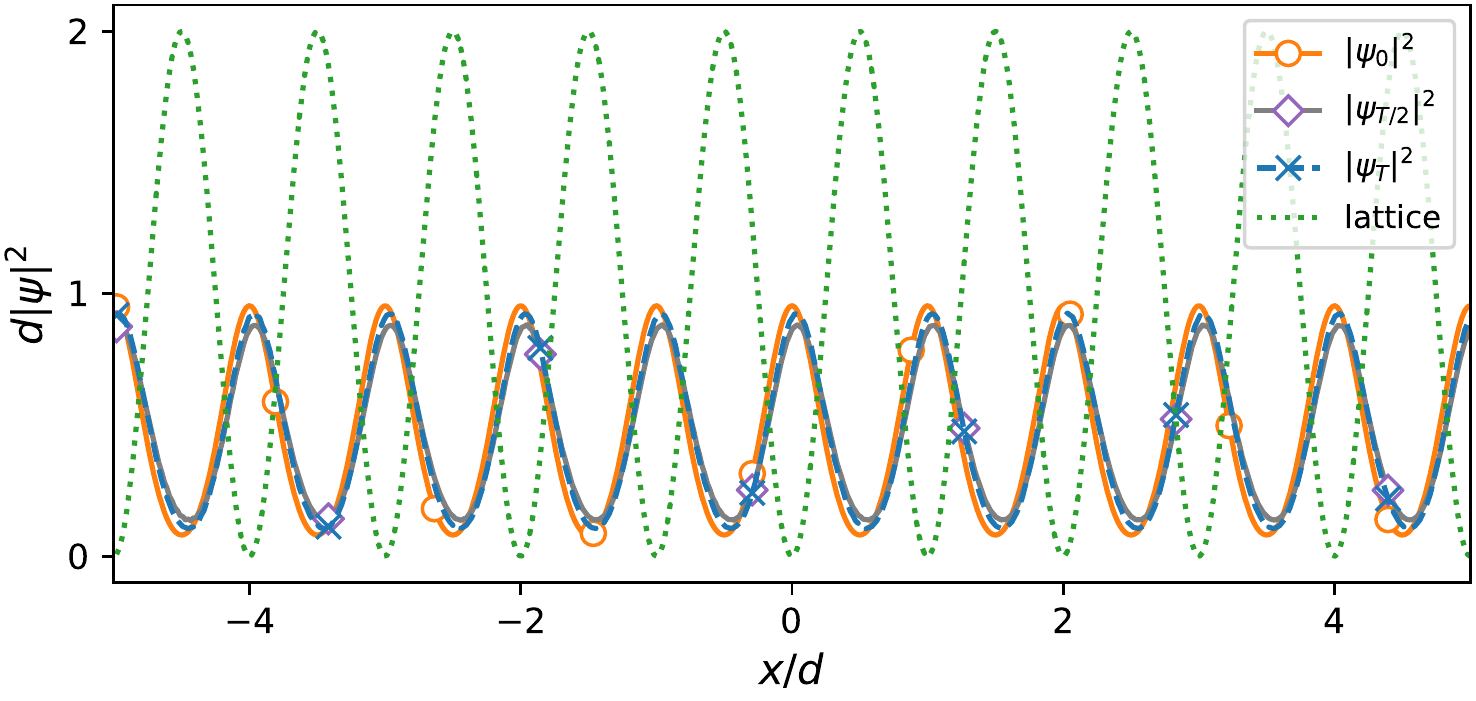}
	\caption{Selected snapshots of the real time evolution of stationary states shown in Fig. \ref{fig:gs} after perturbative noise has been added on the initial states $\psi_0$. In all the cases, the duration of the total time evolution is $T=25\; m d^2/\hbar$. (a) Gap soliton. (b) Nonlinear Bloch states.   } 
	\label{fig:evolution}
\end{figure}
\section{Dynamics}
\label{sec:Dyn}

 Despite the linear analysis performed in the previous section, the stability of stationary states can only be ensured through the nonlinear time evolution of the system. To this end,
we have solved the time-dependent Eq. (\ref{eq:gpe})  for initial stationary states on which perturbative noise has been added. Although our results for the subsequent time evolution are consistent with the predictions  of the linear stability analysis, we have also found some interesting cases whose dynamics show features of structural stability (small variations of the stationary state profile that do not break qualitatively its structure) despite the presence of unstable linear modes.

Figure  \ref{fig:evolution}(a) depicts three snapshots for selected times ($t=0,\,T/2,\,T$) of a time evolution, with total time $T=25 \;md^2/\hbar$, of the linearly-unstable soliton of Fig. \ref{fig:gs}(a) with positive rotation rate.
As can be seen, it confirms the linear prediction of instability shown in Fig. \ref{fig:modes}(b), since it displays the tunneling of particles into nearby lattice sites as time passes. The linear analysis predicts also the instability of the non-linear Bloch states with $k=5k_0$ and opposite rotation rates shown in Fig. \ref{fig:gs}(b), with unstable linear modes of higher energy in the case of positive rotation. However the time evolution, Fig. \ref{fig:evolution}(b), shows a distinct nonlinear dynamics for them. While the positive rotation Bloch state evolves into a non-symmetric structure that is highly variable in time, the negative rotation one exhibits a small breathing dynamics that do not alter the geomteric shape of the initial state. We have checked that different types of small perturbations lead to the same conclusions, and not different dynamics is shown for evolution times much longer (up to ten times) than the case shown in Fig. \ref{fig:modes}(b).

	
	One of the most interesting features of the dynamics in optical lattices is the emergence of Bloch oscillations. This phenomenon has been shown to appear in  BECs with contact interparticle interactions when the lattice velocity is slowly ramped up \cite{Choi1999,Morsch2001}. It manifests the periodic nature of the system through the ground state  transit from positive to negative quasimomentum states over the first energy band. As a consequence, the system's average velocity  oscillates with respect to the lattice velocity.
	But differently to linear systems, the nonlinearity introduced by the interactions in BECs can lead to the breakdown of Bloch oscillations, which is associated with the instability of Bloch states close to the edge of the Brillouin zone \cite{Wu03}. In addition, the energy band structure changes with increasing interactions, so that, precisely at the edge of the Brillouin zone, it develops a cusp, first, and a swallow tail configuration, later, that prevents the adiabatic transit between different quasimomentum states\cite{Wu03}.
	
	In what follows, we make an exploration of Bloch oscillations in systems with current-dependent interactions. On examination of the dispersion of nonlinear Bloch states for varying rotation, Fig. \ref{fig:traj}, and despite the lack of symmetry with respect to the rotation direction, one can expect the transit from positive to negative velocities to take place if the states passed across for varying rotation are dynamically stable, or if, these states being unstable, their instability modes grow at slower rate than the transit speed.
	
	The expected period of Bloch oscillations is
	\begin{equation}
	T_B=\frac{2\pi\hbar}{m \alpha R d},
	\label{eq:TB}
	\end{equation}
	where $\alpha R$ is a constant acceleration in the ring, since this is the time taken to cross the first Brillouin zone from quasi-momentum $k=-\pi/d$ to $k=\pi/d$.  
	 The influence of the nonlinearity on the oscillations in systems with contact interactions can be captured, at least for smooth density profiles, by  effectively modifying the lattice depth as $U_0^{\rm (eff)}=U_0/(1+4{g_{\rm 1D} \bar n}/{E_L})$ \cite{Choi1999}, where $\bar n$ is the average number density. This effective potential provides us with a way to account for the current-density interactions on Bloch oscillations by means of an equivalent effective lattice depth
	\begin{equation}
	U_0^{\rm (eff)}=U_0\,\left(1+4\frac{\hbar\kappa \bar J}{E_L}\right)^{-1},
	\label{eq:veff}
	\end{equation}
  where $\bar J$ is the average current density.

	\begin{figure}[tb]
		\flushleft ({\bf a})\\
		\includegraphics[width=1.0\linewidth]{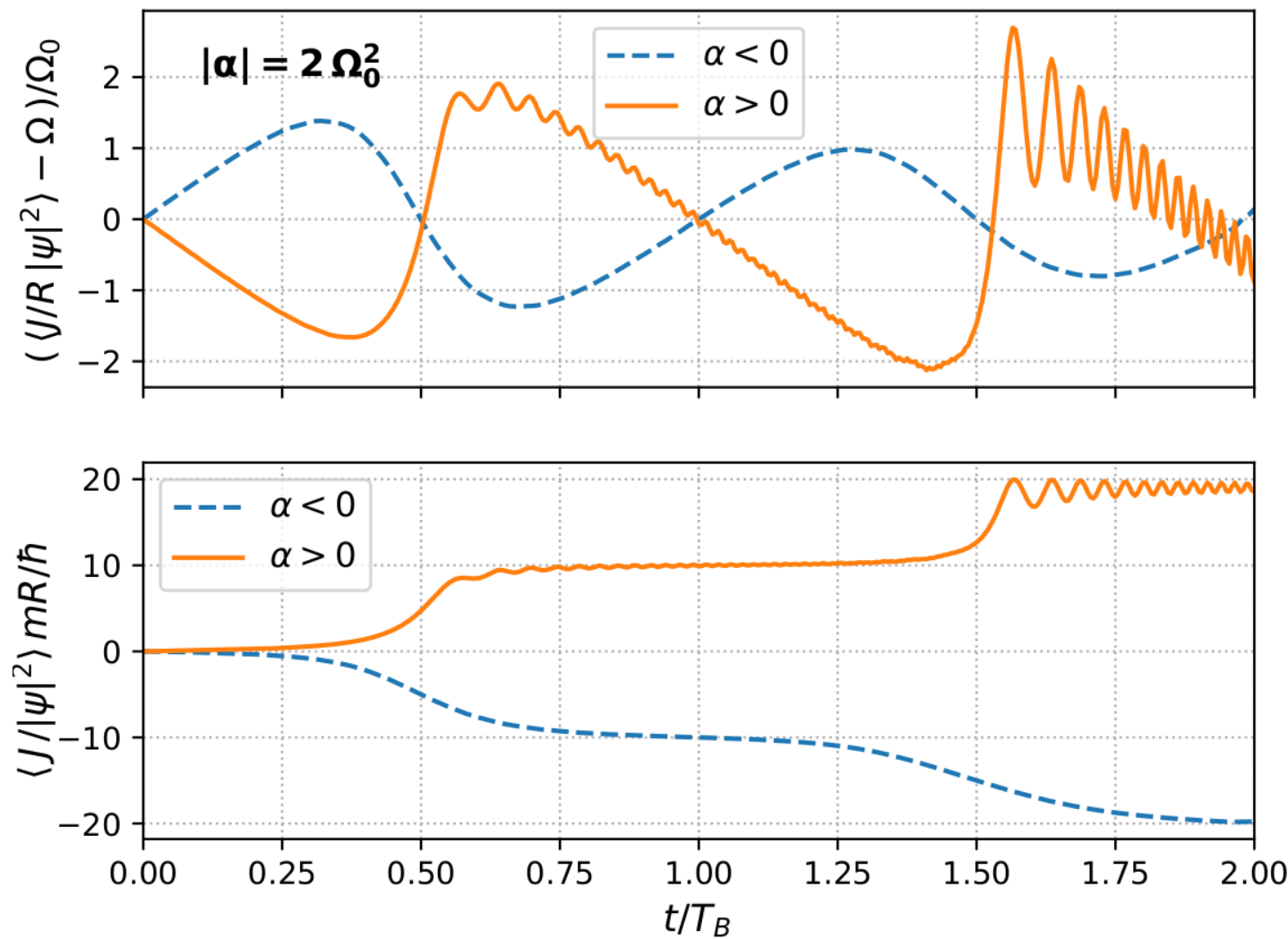}\\
		({\bf b})\\
		\includegraphics[width=1.0\linewidth]{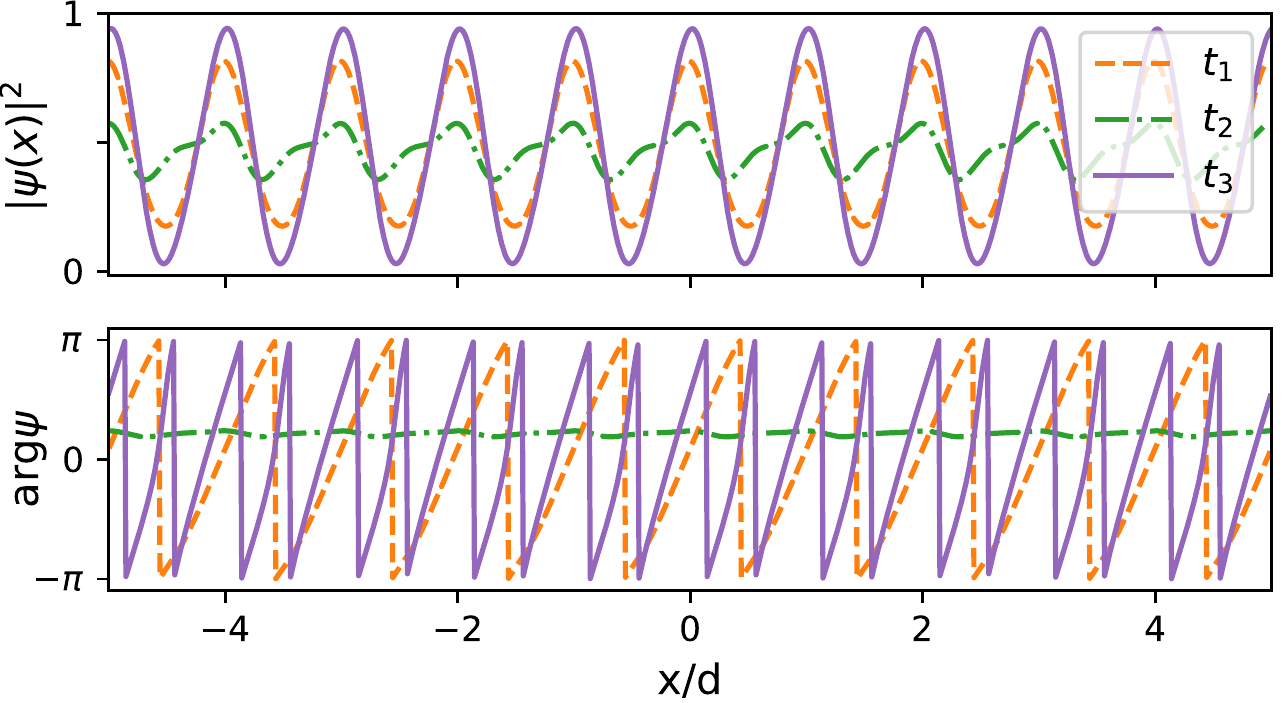}
		
	\caption{Bloch oscillations with current-density interactions. (a) 
	Relative angular rotation (top) and linear velocity (bottom) versus time in a
	lattice with varying rotation rate $\Omega = \alpha t$ for both positive and negative 
	values of the angular acceleration.	
		 (b)  For $\alpha>0$, snapshots of the system state at selected times: $t_1$ at the minimum relative velocity, $t_2$ at the maximum relative velocity, and $t_3$ at the first local minimum after the maximum of the relative velocity. } 
	\label{fig:oscill}
\end{figure}
	
 Guided by Eqs. (\ref{eq:TB}) and (\ref{eq:veff}), we have chosen the initial ground state in Fig. \ref{fig:bands}, with $k=0$, and have imparted a constant angular acceleration $\alpha$ to the ring lattice. The results are shown in Fig. \ref{fig:oscill} for the absolute value of the angular acceleration $|\alpha|/\Omega_0^2=2$. Figure \ref{fig:oscill}(a) represents the relative rotation of the evolved state with respect to the lattice for negative and positive signs of the acceleration.  The observed period of the Bloch oscillations is consistent with Eq. (\ref{eq:TB}), with the zero crossings (or points of zero relative velocity) reached at times $t_j$ that are integer multiples of $T_B/2$, i.e $t_j=j T_B/2$, and $j=0,1,2,\dots$  For $j$ odd,
 the system transits through the edge of the Brillouin zone and
 the wave function presents M nodes (associated with the formation of standing waves by Bragg reflections \cite{Kittel}). The latter fact can be interpreted as
 the presence of M dark solitons caused by the M cores of corresponding
 vortices in their transition from the outer part to the inner part of the ring \cite{Mateo2019}.
 This view is better understood by monitoring the expectation value of the linear velocity
 measured in ring units $\hbar/mR$, as depicted
 in the bottom panel of Fig. \ref{fig:oscill}(a). As can be seen, the linear velocity is approximately 
 constant around integer values of $t/T_B$, that is around $t_j=l\,T_B$, for $j=2l$ even, and  $l=0,1,2,\dots$, when the system transits through the center
 of the Brillouin zone and the velocity takes the value $\sim l M \hbar/mR$, as due to
 $l\times M$ vortices that have entered the inner part of the ring.

  Apparent differences can be observed in the smoothness (and also in the duration, see below) of the Bloch oscillations that depend on the sign of the acceleration. Qualitatively, they can be explained by the effective lattice depth of Eq.~(\ref{eq:veff}). Since positive (respectively negative) rotations translate into positive (resp. negative) current densities, they decrease (resp. increase) the effective lattice depth and 
   make the lattice progressively less (resp. more) relevant. 
 Contrary to the case of contact interactions, this causes the monotonic variation of the amplitude of  Bloch oscillations  (the maximum relative speed) after each zero crossing; it keeps increasing (resp. decreasing) for positive (resp. negative) acceleration.  For high negative values of the current density the superfluid features of the system tend to disappear, the system state becomes strongly localized, and the current density approaches the lattice rotation, which is reflected in the bottom panel of Fig. \ref{fig:oscill}(a) by a progressively straightened curve, without the marked plateaus of the curve with positive acceleration. On the other hand, the latter adiabatic curve becomes  accompanied at higher currents by higher frequency oscillations, related to Landau-Zener tunneling \cite{Choi1999}, between configurations with different density modulation  [see Fig. \ref{fig:oscill}(b)].
 
  We finish by commenting briefly on the effect of different accelerations. As shown in Fig. \ref{fig:oscill}, for $|\alpha|/\Omega_0^2=2$ the decay of Bloch oscillations does not appear before $2T_B$. However, at the lower acceleration of $\alpha=-0.5 \Omega_0^2$ (not shown) the oscillations hardly complete one whole period $T_B$ before breaking down, producing the decay of the homogeneous profile over the lattice sites into more localized density peaks. This decay resembles the action of modulation instabilities \cite{Zakharov2009}. Therefore, our results point to the fact that Bloch oscillations that proceed with (still adiabatic, but) higher acceleration last longer, which we attribute to the fact that in this case the unstable modes have a shorter time to grow.

\section{Conclusions}
\label{sec:Conclusions}

We have reported on gap solitons and nonlinear Bloch states in a rotating ring lattice within a theory with current-density interactions. Our results show that the presence of chirality is manifest in all the states considered, including their spectrum of linear excitations and the display of Bloch oscillations for constant angular acceleration. A novelty is the existence of stationary and dynamically stable non-regular Bloch states characterized by a modulated density profile. 

The recent experimental achievement of this theory in Bose-Einstein condensates of ultracold atoms \cite{Frolian2022} opens the way for the experimental realization of the states and phenomena that we have described. Although currently optical lattice potentials in a ring are experimentally available \cite{Sakamoto2013,Yamane2016}, our results are not restricted to this geometry, and can also be realized in one dimensional linear lattices, as has been routinely done in the presence of contact interatomic interactions \cite{Morsch2006,Schafer2020}.

Future prospects of our work include the study of fundamental and higher-order soliton states in different energy gaps, and the extension to realistic 2D or 3D systems that reach the quasi-1D regime.

\begin{acknowledgments}
JX, QJ, and HQ acknowledge support by the National Natural Science Foundation
of China (Grant No. 11402199), the Natural Science Foundation of Shaanxi Province(Grant No. 2022JM-004, No. 2018JM1050), and the Education Department Foundation of Shaanxi
Province(Grant No. 14JK1676).
\end{acknowledgments}

\bibliography{lattice_dens_dep_gauge.bib}

\end{document}